      [1999/12/01 v1.4c Il Nuovo Cimento]
\documentclass{cimento}

\usepackage{graphicx}  
\usepackage{amsmath}
\title{High energy single top photoproduction at the LHC}
\author{S.~Ovyn\from{ins:x}
\atque J.~de~Favereau~de~Jeneret\from{ins:x}\\
}
\instlist{\inst{ins:x} Universit\'e Catholique de Louvain,\\ Center for Particle Physics and Phenomenology, Louvain-la-Neuve, Belgium}
\PACSes{\PACSit{13.60.-r}{Photon interactions}
\PACSit{14.65.Ha}{Single top production}}
\begin{document}

\maketitle

\begin{abstract}
High-energy photon-proton interactions at the \textsc{lhc} offer interesting possibilities for the study of top properties. Using a fast simulation of a \textsc{lhc}-like detector, first results on the measurement of the $|V_{tb}|$ matrix element using $Wt$ photoproduction are presented. Anomalous photoproduction of single top due to Flavour-Changing Neutral Currents permits to improve the current limits on the coupling parameters, $k_{tu\gamma}$ and $k_{tc\gamma}$ after only 1~fb$^{-1}$.
\end{abstract}

\section{Introduction}

A significant fraction of $pp$ collisions at the \textsc{lhc} will involve (quasi-real) photon interactions occurring at energies well beyond the electroweak energy scale~\cite{bib:piotr}. The \textsc{lhc} can therefore be considered to some extend as a high-energy photon-proton collider. In a recent paper~\cite{bib:nous}, the initial comprehensive studies of high energy photon interactions at the \textsc{lhc} were reported. A large variety of $pp(\gamma g/q \rightarrow X)pY$ processes has sizable cross section and could therefore be studied during the very low and low luminosity phases of \textsc{lhc}. Interestingly, the \textsc{sm} single top quark photoproduction cross section reaches 2.5 pb. Due to the high number of top quarks produced using photon-proton interactions and the large ratio  of single top production cross section to the \textsc{sm} background production cross section, photoproduction offers an ideal framework for studying the top properties like the electric charge and the $|V_{tb}|$ \textsc{ckm} matrix element. Probing the possible anomalous photoproduction of single top via flavour-changing neutral currents (\textsc{fcnc}) is also favoured at the \textsc{lhc} because of the high expected cross section compared to the one at \textsc{hera}. For same values of the anomalous coupling $k_{tu\gamma}$, the cross section is expected to be two orders of magnitude higher. Furthermore, while at \textsc{hera} only the up quark content of the proton contributed, the energy of the \textsc{lhc} allows to probe the proton at lower momentum fraction, opening the opportunity to probe the effect of the c quark via the $k_{tc\gamma}$ coupling.

\section{Events simulation}

All samples used in the analysis of \textsc{sm} and anomalous single top photoproduction have been produced using the adapted MadGraph/MadEvent~\cite{bib:mgme1,bib:mgme2} and Calc\textsc{hep}~\cite{bib:calchep} programs. In order to take into account the effect of jet algorithms and the efficiency of event selection under realistic experimental conditions, the generated events were passed: (1) to \textsc{pythia} 6.227~\cite{bib:pythia} and (2) a fast simulation of a typical \textsc{lhc} multipurpose detector. This simulation assumes geometrical acceptance of sub-detectors and their finite energy resolutions. More information can be found in reference~\cite{bib:nous}.

\section{Suppression of $pp$ interactions}
\label{sec:ppsupp}

Tagging is essential for the extraction of high energy photon-induced interactions from the huge $pp$ events. Photon-induced interactions are characterised by a large pseudorapidity region completely devoid of any hadronic activity. This region is usually called {\it large rapidity gap} (\textsc{lrg}).

\subsection{Very low luminosity phase ($< 10^{33}$ cm$^{-2}$s$^{-1}$)}

The number of extra interactions per beam crossing (\textit{pile-up}) is negligible. Thanks to the colour flow in $pp$ interactions between the proton remnant and the hard hadronic final states, a simple way to suppress generic $pp$ interactions is to require \textsc{lrg}s by looking at the energy measured in the forward detector containing the minimum forward activity ($3<|\eta|<5$). For a maximal allowed energy of 50~GeV, a typical reduction factor of 10$^{-3}$ and 10$^{-2}$ for a parton-parton $t\overline{t}$ and $Wj$ production respectively is expected. This cut, denoted as $E^{FCal}$, can be done using the central detector only, however the event kinematics is less constrained. A total integrated luminosity of 1~fb$^{-1}$ for such no {\it pile-up} condition seems to be a realistic assumption. The rejection can be further improved by using an exclusivity condition requiring no additionnal tracks ({\it i.e.} excluding isolated leptons and jet cones) with $p_T>$~0.5~GeV and 1~$<\eta<$~2.5 in the hemisphere where the rapidity gap is present. With these defined acceptance cuts, rapidity gap and exclusivity conditions, efficiency for signal processes drops roughly by a factor of two while the reduction factors for parton-parton reactions are better than $10^{-3}$. Providing good control of the energy scale of forward calorimeters and efficient tagging based on \textsc{lrg}s, one expects inclusive parton-parton processes to be of the same order of magnitude than the irreducible photon-induced backgrounds.

\subsection{Low luminosity phase}

The \textsc{lrg} technique cannot be used because of large event {\it pile-up}. The exclusivity condition alone cannot reduce partonic backgrounds to a level that allows proper signal extraction. In that case,  in addition to the exclusivity condition, the use of \textit{very forward detectors} (\textsc{vfd}) to detect the escaping proton is mandatory in order to retain $pp$ background low. However, \textsc{vfd}s cannot provide a total rejection of the partonic processes because of the presence of single diffractive events in the {\it pile-up}. Hence, the overall event mimics well a photoproduction event. The probability of such accidental coincidences provides directly the rejection power of \textsc{vfd}s. For instance, the case for which \textsc{vfd} stations would be put at 220~m and 420~m from the interaction point has been computed and provides rejection factors of 11 and 5.6 for $10^{33}~\textrm{cm}^{-2}~\textrm{s}^{-1}$ and $2 \times 10^{33}~\textrm{cm}^{-2}~\textrm{s}^{-1}$ luminosity respectively.

\section{Backgrounds}

Backgrounds considered in this contribution come from photon-proton and proton-proton interactions. Inelastic photoproduction, in which the proton having emitted a photon does not survive the interaction, has not been taken into account. This would add to the cross section of both signal and photon-induced backgrounds. The cross sections for such inelastic processes is not precisely known, because the probability of rescattering becomes important due to the small impact parameter. This makes the efficiency of tagging such events much harder to compute precisely, leading to important systematic errors. Diffractive background processes have also not been considered here, although they can look very similar to photoproduction.

\section{W associated single top photoproduction}

\begin{figure}[!h]
\begin{center}
\includegraphics[width=.5\linewidth]{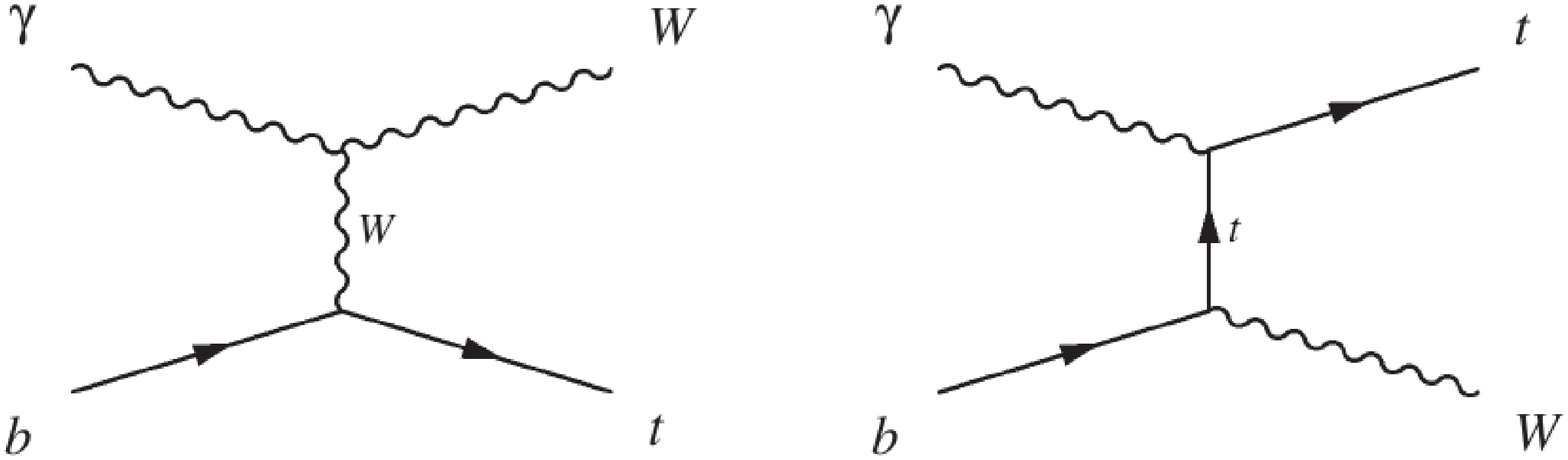}
\caption{Diagrams for the dominant contribution to the \textsc{sm} production of single top quark.}
\label{fig:ovyn_fig1}
\end{center}
\end{figure}

Photoproduction of single top is dominated at tree-level by t-channel amplitudes when the top quark is produced in association with a $W$ boson (fig.~\ref{fig:ovyn_fig1}). In contrast to proton-proton interactions where the ratio of $Wt$ associated production cross section to the sum of all top production cross sections is only about $5\%$, it is about 10 times higher in photoproduction. This provides a unique opportunity to study this reaction at the start phase of the \textsc{lhc}. While the overall photoproduction of top quark is sensitive to the top quark electrical charge, the $Wt$ associated photoproduction amplitudes are all proportional to the \textsc{ckm} matrix element $|V_{tb}|$.

The $pp(\gamma q \rightarrow Wt)pY$ process results in a final state of two on-shell $W$ bosons and a $b$ quark. The studied topologies are $\ell bjj$ for the semi-leptonic decay of the two $W$ bosons and $\ell \ell b$ for the di-leptonic decay, where $\ell=e,~\mu~\textrm{or}~\tau$. The cross sections of these final states are 440~fb and 104~fb respectively. The dominant irreducible background of both channels is expected to stem from the $t\overline{t}$ production, where a jet misses the acceptance region. Other backgrounds are $Wb\bar{b}q'$, $Wjjj$ and $WWq'$ processes produced either from photon-proton interactions, either from proton-proton interactions. Their cross sections including the branching ratio in the desire topology are summarised in table~\ref{tab:ovyn_tab1}.

\begin{table}[!h]
\renewcommand{\arraystretch}{1.2}
\caption{Background processes used in the semi-leptonic and di-leptonic channels. Cross-sections include generation cuts of $p_T>$~1~GeV for $q'$ and $p_T>$~10~GeV for jets. Branching ratio of the W boson into leptons (e, $\mu$ or $\tau$) is taken into account.}
\begin{tabular}{lrrrr}
\hline
Final state & \multicolumn{2}{c}{Semi-leptonic} & \multicolumn{2}{c}{Di-leptonic}\\
            & $\sigma$ [fb] & sample size & $\sigma$ [fb] & sample size\\
\hline
$\gamma p \rightarrow t\bar{t}$            & 672~~~              & 170~k  & 159~~~            & 100~k       \\
$~~~~~~~~Wjjj$                             & 2.8  $\times 10^3$  &  50~k  & -~~~~~            & -~~~     \\
$~~~~~~~~W b\bar{b} q'$                    & 55~~~~              &  50~k  & -~~~~~            & -~~~     \\
$~~~~~~~~WWq'$                             & -~~~~~              & -~~~   & 63~~~~            & 70~k\\\hline
$pp \rightarrow t\bar{t}$                  & 329  $\times 10^3$  & 390~k  & 78 $\times 10^3$  & 130~k      \\
$~~~~~~~~W jets$                      & 73  $\times 10^6$   & 770~k  & -~~~~~            & -~~~     \\
$~~~~~~~~W b\bar{b} j$                     & 267  $\times 10^3$  & 120~k  & -~~~~~            & -~~~      \\
$~~~~~~~~tj$                               & 67  $\times 10^3$   & 100~k  & -~~~~~            & -~~~     \\ 
$~~~~~~~~WWj$                              & -~~~~~              & -~~~   & 5.2 $\times 10^3$ & 50~k\\
\hline
\end{tabular}
\label{tab:ovyn_tab1}
\end{table}

\subsection{Signal selection}

For the semi-leptonic final state, the topology-based cuts applied are the following: one isolated lepton with $p_T^{\ell}>20$~GeV; exactly 3 jets with $p_T^{j}>30$~GeV. To ensure that partonic backgrounds are reduced to the same level as photoproduction ones, the $E^{FCal}$ selection with a cut value of 30~GeV as well as the exclusivity cut are applied. Exactly one of the three jets must be b-labbeled. Two cuts based on kinematics are also applied: the invariant mass of the two non b-tagged jets must satisfy $|m_Z-m_{jj}|<20$~GeV and the scalar sum of the visible objects must be smaller than 230~GeV. After this selection, the final cross section for the signal is reduced to 4.8~fb, against 5.5~fb for the backgrounds, 65~$\%$ of which comes from partonic processes. The $t\bar{t}\rightarrow \ell \ell b\bar{b}$ topology is also taken into account in the backgrounds. Details are given in table~\ref{tab:ovyn_tab2}.

The procedure to select the di-leptonic topology is simpler: two isolated lepton with $p_T>20$~GeV; one b-tagged jet with $p_T^{b}>30$~GeV; no additional jets with $p_T^j>30$~GeV and the presence of transverse missing energy higher than 20~GeV. The same rapidity gap and exclusivity condition as in the semi-leptonic topology are applied. Signal cross section for this topology is 4.9~fb after cuts, for a background cross section of 2.2~fb with less than 30~$\%$ of partonic contribution. Details are in table~\ref{tab:ovyn_tab2}.

\begin{table}[!h]
\caption{Effect of various cuts on the cross section of the semi-leptonic and di-leptonic topologies, photoproduction backgrounds and partonic backgrounds.}
\renewcommand{\arraystretch}{1.2}
\begin{tabular}{lcccccc}
\hline
$\sigma$ [fb] & \multicolumn{3}{c}{Semi-leptonic topology} & \multicolumn{3}{c}{Di-leptonic topology}\\
              & signal & $\gamma p$ & $pp$ & signal & $\gamma p$ & $pp$\\
\hline
production    &  440.0 & 3.6 $\times 10^3$ & 74 $\times 10^6$  & 104.0 & 222   & 83 $\times 10^3$\\
topology cuts &  36.0  & 144.4             & 116 $\times 10^3$ & 14.2  & 13.7  & 3.4 $\times 10^3$\\
gap + exclu.  &  24.2  & 77.9              & 187.5             & 12.7  & 8.0   & 3.2                  \\
final cuts    &  4.8   & 1.9               & 3.6               & 4.9   & 1.6   & 0.7              \\\hline
Expected yield for 10 fb$^{-1}$ & 48 & 19 & 36 & 49 & 16 & 7\\
\hline
\end{tabular}
\label{tab:ovyn_tab2}
\end{table}

\subsection{Systematic errors}

When no estimate on the theoretical uncertainties is found in the literature for photon-proton events, the same uncertainty as for the corresponding partonic process is taken for a pessimistic estimate. Partonic cross sections after cuts are considered known to the 2~$\%$ level as the cross section without application of the $E^{FCal}$ and exclusivity conditions can be measured directly and the error on the effect of these cuts is computed separately. The most relevant detector systematics are expected to be the uncertainties on the Jet Energy Scale (\textsc{jes}), on the number of tracks reconstructed in order to apply the exclusivity condition and on the energy measurement in the forward calorimeter. The uncertainty due to \textsc{jes} is expected to be $5\%$ for jets with $p_T < 30~\textrm{GeV}$, $3\%$ for jets with $p_T > 50~\textrm{GeV}$ and a linear interpolation between these two boundaries. The systematic uncertainty due to the exclusivity condition is estimated by moving the track reconstruction efficiency, fixed to 90 $\%$ by default, to 85 $\%$ and 95~$\%$. Finally, the cut on the energy in the forward calorimeter of the gap side has been moved by 10~$\%$ upwards and downwards in order to have an idea of the $E^{FCal}$ uncertainty. The b-tagging uncertainty is taken as $\pm5\%$, while no error on mis-tagging is assumed. The uncertainty due to luminosity is expected to be $5\%$. Using the expected yields in table~\ref{tab:ovyn_tab2}, we can estimate the effect of these uncertainties (table~\ref{tab:ovyn_tab3}). The total error is dominated by the rapidity gap and exclusivity cut errors.

\begin{table}
 \caption{Systematic errors on signal and backgrounds for the semi-leptonic and di-leptonic topologies.}
 \renewcommand{\arraystretch}{1.2}
 \begin{tabular}{lcccc}
  \hline
Error            & \multicolumn{2}{c}{di-leptonic topology} & \multicolumn{2}{c}{semi-leptonic topology}\\
                 & signal ($\%$) & background ($\%$) & signal ($\%$) & background ($\%$)\\
\hline
\textsc{jes}     & 0.6           & 3.7  &      6.7     &    10.6\\
Rapidity gap     & 0.1           & 10.8 &      0.5     &    10.5\\
Exclusivity      & 0.4           & 6.6  &      1.2     &    6.2\\
b-tagging        & 5.0           & 0.0  &      5.0     &    0.0\\
Theoretical      & 6.0           & 3.4  &      6.0     &    2.0\\
Luminosity       & 5.0           & 5.0  &      5.0     &    5.0\\
\hline
 \end{tabular}
\label{tab:ovyn_tab3}
\end{table}

\subsection{Results}

The cross section is calculate as:

\begin{equation*}
\sigma = \frac{S}{\varepsilon L} = \frac{N-B}{\varepsilon L},
\end{equation*}

where S is the number of observed events, N the total number of observed events, B the expected number of background events, $\varepsilon$ the estimated efficiency and $L$ the integrated luminosity. A simple propagation of errors shows that the relative uncertainty on the measured cross section is given by the following formula :

\begin{equation*}
\frac{\Delta \sigma_{obs}}{\sigma_{obs}} = \frac{\Delta \varepsilon}{\varepsilon} \oplus \frac{\Delta L}{L} \oplus \left[ \frac{B}{S} \right] \frac{\Delta B}{B} \oplus \left[ \frac{B}{S} + 1 \right] \frac{\Delta N}{N},
\end{equation*}

where $\Delta \varepsilon$, $\Delta L$ and $\Delta B$ are the systematic errors estimates on the signal selection efficiency, the luminosity and the background cross section respectively and $\Delta N$ is the statistical error on the observed number of events. After an integrated luminosity of 10~fb$^{-1}$, the following uncertainties are obtained for the di-leptonic and the semi-leptonic topologies respectively:

\begin{eqnarray*}
\textrm{Di-leptonic topology: } & \frac{\Delta \sigma_{obs}}{\sigma_{obs}} & = 5.3 \oplus 5.0 \oplus 6.4  \oplus 17.4 = 19.4 \%\\
\textrm{Semi-leptonic topology: } &  \frac{\Delta \sigma_{obs}}{\sigma_{obs}} & = 8.5 \oplus 5.0 \oplus 17.3 \oplus 17.9 = 33.3 \%\\
\end{eqnarray*}

From the error on the total single top cross section measurement, one can compute the error on the $|V_{tb}|$ measurement from the formula $\frac{\Delta |V_{tb}|}{|V_{tb}|} = \frac{1}{2} \left[ \frac{\Delta \sigma_{obs.}}{\sigma} \oplus \frac{\Delta \sigma_{theo.}}{\sigma} \right]$. The expected error on the measurement of $|V_{tb}|$ is 16.9 $\%$ for the semi-leptonic channel and 10.1 $\%$ for the leptonic one after 10~fb$^{-1}$ of integrated luminosity, while the expected uncertainty from the equivalent study based on partonic interactions is 14 $\%$ \cite{bib:cmsvtb} using the same integrated luminosity, showing that photoproduction is at least competitive with partonic-based studies and that the combination of both studies could lead to significant improvement of the error.

\section{Anomalous single top photoproduction}

\textsc{fcnc} appear in many extensions of the Standard Model, such as two Higgs-doublet models or R-Parity violating supersymmetry. The observation of a large number of single top events at the \textsc{lhc} would hence be a clean signature of \textsc{fcnc} induced by processes beyond the Standard Model. The effective Lagrangian for this anomalous coupling can be written as~\cite{bib:eff_lag_anotop}:

\begin{equation*}
\mathcal{L} =  iee_t\bar{t}~\frac{\sigma_{\mu\nu}q^{\nu}}{\Lambda}~k_{tu\gamma}uA^{\mu}
 + iee_t\bar{t}~\frac{\sigma_{\mu\nu}q^{\nu}}{\Lambda}~k_{tc\gamma}cA^{\mu} + h.c.,
\end{equation*}
where $\sigma^{\mu\nu}$ is defined as $(\gamma^{\mu} \gamma^{\nu} - \gamma^{\nu} \gamma^{\mu})/2$, $q^{\nu}$ being the photon 4-vector and $\Lambda$ an arbitrary scale, conventionally taken as the top mass. The couplings k$_{tu\gamma}$ and k$_{tc\gamma}$ are real and positive such that the cross section takes the form $\sigma_{pp \rightarrow t} = \alpha_u\ k^2_{tu\gamma} + \alpha_c\ k^2_{tc\gamma}$. The computed $\alpha$ parameters obtained using \textsc{c}alc\textsc{hep} are $\alpha_u = 368$~pb and $\alpha_c = 122$~pb. The present upper limit on $k_{tu\gamma}$ is around 0.14, depending on the top mass~\cite{bib:h1_ktug} while the anomalous coupling $k_{tc\gamma}$ has not been probed yet. The studied final state consists of a leptonic decay of the W boson coming from the top quark, giving a final topology consisting of a hard lepton and a jet from the b quark. The dominant background processes for this final state come from events with one W boson and one jet mis-tagged as a b-jet. The contribution of genuine b-jets is negligible because of the low cross section of this process: three orders of magnitude lower than the cross section of the $Wc$ topology. Backgrounds cross sections and sample sizes are given in table~\ref{tab:ovyn_tab4}. 

\begin{table}[!h]
\renewcommand{\arraystretch}{1.2}
\caption{Background processes used for the analysis of the anomalous top photoproduction. Cross-sections include the branching ratio of the W boson to electron or muon and generation cuts of $p_T~>~10~\textrm{GeV}$ for leptons and $p_T~>~20~\textrm{GeV}$ for jets.}
\begin{tabular}{lcrcr}
\hline
Process & \multicolumn{2}{c}{$\gamma p$ events} & \multicolumn{2}{c}{$pp$ events}\\
& $\sigma$ [fb] & sample size & $\sigma$ [fb] & sample size\\
\hline
$W j$  & 41.6 $\times 10^3$ & 100 k  &  77.3 $\times 10^6$ & 100 k   \\
$W c$  & 11.5 $\times 10^3$ & 100 k  &  8.8 $\times 10^6$ & 100 k   \\
\hline
\end{tabular}
\label{tab:ovyn_tab4}
\end{table}

\subsection{Signal selection}

Preselection cuts require the presence of exactly one jet with $p_T~>~45$~GeV and one isolated lepton with $p_T~>~20$~GeV. The two cuts designed to reject $pp$ interactions in the scheme of zero pile-up conditions are applied with a maximum allowed energy in the forward hemisphere of 20~GeV. An event is selected if the only allowed jet is tagged as a b-jet. This requirement has a large rejection power against processes that do not contain a true b-jet in the final state. A top candidate is also reconstructed from the W-boson and the ``b-jet". 


In order to extend this study in presence of {\it pile-up}, the use of the $E^{FCal}$ selection cut is replaced by the tagging of the escaping proton by \textsc{vfd}s as described in sec.~\ref{sec:ppsupp}. A proper simulation of the proton propagation in the \textsc{lhc} beamline performed using \textsc{hector}~\cite{bib:hector}, shows that using detectors stations at 220~m and 420~m from the \textsc{ip}, one selects events for which the proton has lost between 20~GeV and 800~GeV. As stated before, the reduction of the partonic background is not strong enough to reach the same level as the photoproduction backgrounds. Another advantage of the \textsc{vfd} is that, considering a well designed reconstruction algorithm, the energy loss of the proton that hits the detector can be determined and used to improve the selection of photoproduction processes. An additional cut is therefore used that reconstructs the top quark longitudinal momentum both from the central event and from the proton energy loss. The difference between these two values allows to distinguish between photoproduction events for which they are close, and partonic events for which the distance between them is distributed randomly.

\subsection{Systematic errors}

The same systematic uncertainties as in the case of the \textsc{sm} single top study have been estimated. One again, the rapidity gap and exclusivity condition account for the most important part of it. In the case of higher luminosities for which the \textsc{vfd}s where used, no systematic error is assumed on this tagging. The detail of all errors for both scenario's are given on table~\ref{tab:ovyn_tab5}. Signal systematics stay unaffected by the scenario change, as the error due to the \textsc{lrg} requirement is negligible.

\begin{table}[!h]
 \caption{Systematic errors on signal and backgrounds for both scenarios.}
 \renewcommand{\arraystretch}{1.2}
 \begin{tabular}{lccc}
  \hline
Error         & signal ($\%$) & \multicolumn{2}{c}{Background ($\%$)}\\
              &               & very low          & low              \\
\hline
\textsc{jes}  & 1.6           & 3.0               & 3.3 \\
\textsc{lrg}  & 0.0           & 9.9               &  -  \\
Exclusivity   & 1.0           & 5.5               & 6.9 \\
Luminosity    & 5.0           & 5.0               & 5.0 \\
Theoretical   & 5.0           & 1.9               & 1.3 \\
b-tagging     & 5.0           & 0.0               & 0.0 \\
\hline 
 \end{tabular}
 \label{tab:ovyn_tab5}
\end{table}

\subsection{Results}

Using the \textsc{lrg} requirement for an integrated luminosity of 1~fb$^{-1}$ and the \textsc{vfd}s for an integrated luminosity of 30~fb$^{-1}$, one gets the following final result ($k_{tu\gamma} = 0.15$, $k_{tc\gamma} = 0$) : 

\begin{align*}
\textrm{Very~low~luminosity} \left\{
\renewcommand{\arraystretch}{1.2} 
\begin{array}{rr}
\textrm{Signal:}& 83.2 \pm 9.1~ \textrm{(stat.)}~ \pm 7.4~ \textrm{(syst.)~events}\\
\textrm{Background:}& 12.7 \pm 3.6~ \textrm{(stat.)}~ \pm 1.6~ \textrm{(syst.)~events} \end{array}\right.\\
\textrm{Low~luminosity} \left\{
\renewcommand{\arraystretch}{1.2} 
\begin{array}{rr}
\textrm{Signal:}& 1554 \pm 39~ \textrm{(stat.)}~ \pm 138~ \textrm{(syst.)~events}\\
\textrm{Background:}& ~327 \pm 18~ \textrm{(stat.)}~ \pm 30~ \textrm{(syst.)~events} \end{array}\right.
\end{align*}

In order to set a limit on the anomalous couplings, we assume a measurement in agreement with the \textsc{sm}, {\it i.e.} seeing the background only. Given this measurement, the maximum cross section for which this measurement is not less than 5~$\%$ probable is computed. This cross section corresponds to the minimum real cross section for which the ``\textsc{sm} only'' hypothesis will be rejected at 95~$\%$ C.L. and thus gives the minimum anomalous cross section once the \textsc{sm} cross section is subtracted. The number of events is assumed to be distributed according to a Poisson distribution, while the systematic uncertainty is included using Monte Carlo to obtain a realistic convolution of statistical and systematic errors. The obtained expected limits on the anomalous couplings are the following : 

\begin{eqnarray*}
\textrm{Very low luminosity : } k_{tu\gamma} < 0.044,~ k_{tc\gamma} < 0.077, \\
\textrm{Low luminosity : } k_{tu\gamma} < 0.029,~ k_{tc\gamma} < 0.050. \\
\end{eqnarray*}

\section{Conclusions and prospects}

Both above studies show the potential to bring precise measurement of important parameters linked to top quark physics. The $|V_{tb}|$ measurement with a precision of 10~$\%$ to 17~$\%$ should be competitive to the partonic processes-based one, allowing to improve it. The anomalous couplings limits should be at least 3-4 times better than the ones obtained at \textsc{hera}. However, both studies will be refined when full detector simulation will be used, providing a better estimate of the systematic errors in order to replace the pessimistic ones used in the present analysis. Also, studying the influence of diffractive backgrounds is an important part of the work to be done, as well as the contribution of inelastic photon emissions to both signal and backgrounds.


\end{document}